\documentclass[pra,superscriptaddress,floatfix,showpacs,aps,reprint,letterpaper,nobalancelastpage,footinbib,longbibliography]{revtex4-1}

\usepackage{amsmath,amsfonts,amssymb,amsthm,mathtools}
\usepackage[final]{graphicx}
\usepackage[dvipsnames]{xcolor}
\usepackage[urlcolor=blue, hyperindex, colorlinks, bookmarks=true]{hyperref}
\hypersetup{
  linkcolor=red,
  citecolor=blue
}
\usepackage{qip}

\makeatletter \renewcommand\d[1]{\ensuremath{%
  \;\mathrm{d}#1\@ifnextchar\d{\!}{}}}
\makeatother

\usepackage{microtype}

\begin{document}

\title{Experimental Pauli-frame randomization on a superconducting qubit}

\date{\today}

\author{Matthew Ware}
\affiliation{Raytheon BBN Technologies, 10 Moulton St., Cambridge, MA
  02138, USA}

\author{Guilhem Ribeill}
\affiliation{Raytheon BBN Technologies, 10 Moulton St., Cambridge, MA
  02138, USA}

\author{Diego Rist\`{e}}
\thanks{Current address: Quantum Engineering Solutions, Keysight
  Technologies, One Broadway, Cambridge, MA 02412}

\author{Colm A. Ryan}
\thanks{Current address: AWS Center for Quantum Computing, Pasadena,
  CA 91125}

\author{Blake Johnson}
\thanks{Current address: IBM T.J. Watson Research Center, Yorktown
  Heights, NY 10598, USA; blake.johnson@ibm.com}

\author{Marcus P. da Silva}
\thanks{Current address: Microsoft Quantum, One Microsoft Way,
  Redmond, WA 98052; marcus.silva@microsoft.com}

\begin{abstract}
  The promise of quantum computing with imperfect qubits relies on the
  ability of a quantum computing system to scale cheaply through error
  correction and fault-tolerance. While fault-tolerance requires
  relatively mild assumptions about the nature of qubit errors, the
  overhead associated with coherent and non-Markovian errors can be
  orders of magnitude larger than the overhead associated with purely
  stochastic Markovian errors.  One proposal to address this challenge
  is to randomize the circuits of interest, shaping the errors to be
  stochastic Pauli errors but leaving the aggregate computation
  unaffected. The randomization technique can also suppress couplings
  to slow degrees of freedom associated with non-Markovian evolution.
  Here we demonstrate the implementation of {\em Pauli-frame
    randomization} in a superconducting circuit system, exploiting a
  flexible programming and control infrastructure to achieve this with
  low effort.  We use high-accuracy gate-set tomography to
  characterize in detail the properties of the circuit error, with and
  without the randomization procedure, which allows us to make
  rigorous statements about Markovianity as well as the nature of the
  observed errors.  We demonstrate that randomization suppresses
  signatures of non-Markovian evolution to statistically insignificant
  levels, from a Markovian model violation ranging from $43\sigma$ to
  $1987\sigma$, down to violations between $0.3\sigma$ and $2.7\sigma$
  under randomization.  Moreover, we demonstrate that, under
  randomization, the experimental errors are well described by a Pauli
  error model, with model violations that are similarly insignificant
  (between $0.8\sigma$ and $2.7\sigma$).  Importantly, all these
  improvements in the model accuracy were obtained without degradation
  to fidelity, and with some improvements to error rates as quantified
  by the diamond norm. This demonstrates the ability of Pauli-frame
  randomization to shape noise into forms that are more benign for
  quantum error correction and fault-tolerance.
\end{abstract}

\maketitle

\section{Introduction} \label{sec:introduction}
\setcounter{secnumdepth}{1}

Large-scale quantum computation poses a number of design and control
challenges.  Significant efforts are in progress \cite{KBF+17,TCC+17,
  CGM+14} to meet and overcome challenges associated with initial
state preparation, maintaining coherence, implementing universal
gates, and measuring qubits reliably -- all key criteria for building
scalable quantum computers~\cite{DiV00}.  As the system coherence
times continue to grow, coherent errors can become the dominant source
of error. These errors can originate from miscalibration of qubit
rotations, unintentional control frequency detunings, or interactions
between systems that are otherwise assumed to be decoupled---all
ubiquitous problems for experimental quantum computers. These errors
are also particularly difficult to simulate in multiqubit systems, as
they can interfere constructively and destructively, making prediction
about the performance of quantum error correction codes and
fault-tolerant computation quite difficult~\cite{ISP+16, Pou+17,
  IP17}. Moreover, theoretical lower bounds on the tolerable rates for
coherent errors indicate they may be much more damaging than
stochastic errors~\cite{TB05,AGP06,AB08,NP09}. One way to address this
problem is to transform coherent errors into incoherent, stochastic
errors, such as random bit and phase flips.  Here we use a
superconducting qubit system to implement {\em Pauli-Frame
  Randomization} (PFR)~\cite{KAS05,K05,Knill05,WE16} and show that
coherent errors can be reshaped into stochastic Pauli errors. We also
discuss some additional benefits of the randomization process, such as
decoupling of slow non-Markovian noise~\cite{VK05}.

One significant challenge in determining whether PFR has indeed made
coherent errors stochastic is their small magnitude. Thanks to the
community's progress towards fault-tolerance, the magnitudes of these
errors are on the order of $10^{-3}$ or less in state-of-the-art
devices.  Measuring such small errors reliably runs into limitations
of various characterization approaches: standard tomography is
sensitive to preparation and measurement imperfections and has very
low accuracy, while randomized benchmarking estimates a quantity
(closely associated with the average
fidelity~\cite{Nielsen02,PRY+17,W+17}) that does not differentiate
between coherent and stochastic errors, and cannot test if errors
corresponds to Pauli error models or not. In this demonstration we use
{\em gate set
  tomography}~(GST)~\cite{Sta14,MGS+13,BK+15,BGN+17,NGR+20}, a
tomographic reconstruction technique that provides 1) insensitivity to
state preparation and measurement (SPAM) errors 2) nearly
quantum-limited accuracy 3) an open source library for experiment
design and data analysis~\cite{PYGSTI}. Critically, GST also allows us
to accurately quantify not only the behavior of the diamond norm
error~\cite{KSV02,W+09} and average infidelity~\cite{Nielsen02} under
randomization, but also detailed features of individual gate errors
and the degree to which the evolution is well described by a
Markovian, time-invariant model~\footnote{ We define the operations to
  be Markovian if they can be well approximated by linear
  transformations of the state (density operator)---we
  consider the coarse-grained evolution at the time scale of the
  primitive gates, and do not consider issues such as dividibility of
  operations, or their infinitesimal generators.}---all of which help
confirm the predicted Pauli error model behavior, despite the presence
of general imperfections in the randomization operations.

The remainder of the paper is organized as follows.  In
Section~\ref{sec:frame-rand} we describe PFR, and discuss how to test
its implementation, in a statistically rigorous manner, in
Section~\ref{sec:CV}.  Section~\ref{sec:experiments} describes the
experiments as well as the infrastructure required to create and
process randomized sequences.  Finally, in Section~\ref{sec:results}
we discuss the experimental results.

\section{Pauli-frame randomization} \label{sec:frame-rand}

Pauli-frame randomization (PFR) is a noise-shaping technique that
reduces general noise to effective random Pauli errors between
computational gates~\cite{KAS05,K05,Knill05,WE16}. If the
computational gates consist of Clifford group operations~\cite{GC+99}
(a set of operations sufficient for the most promising approaches to
error correction and fault-tolerance), the effect of these random
Pauli operations can be easily tracked~\cite{Got99,AG04} so that the
computation can be unrandomized by simply reinterpreting the
measurement results. While this randomization is designed to have no
impact on the ideal computation, it effectively {\em symmetrizes} the
error, much like {\em
  twirling}~\cite{BDSW96,DLT02,ESM+07,DCEL09,SMKE08,MSRL12} and
randomized decoupling~\cite{VK05}, leading to an effective error
operation that corresponds to a mixture of Pauli group operations
known as a {\em Pauli channel}, or a {\em Pauli error model}.

These results can be derived in the limit of perfect randomization
operations and gate-independent errors as follows. Consider a set of
ideal (resp. noisy)~\footnote{Throughout this discussion, we represent
  quantum operations by the corresponding superoperators, denoted with
  calligraphic upper-case letters ($\sop{A}$, $\sop{B}$, etc.), while
  the noisy implementations are denoted in the same way but with a
  tilde.} Clifford group quantum operations $\sop{C}_i$
(resp. $\tilde{\sop{C}}_i$). Any sequence of ideal Clifford operations
can be randomized by inserting uniformly random Pauli group operations
between the Clifford group operations. Since Clifford group operations
transform Pauli group operation to other Pauli group operations, the
overall effect of these random Pauli group operations can be cancelled
out by applying a final single Pauli group operation at the end of the
sequence of gates. Moreover, since the Pauli group is a subgroup of
the Clifford group, one may simply combine the $i$th random Pauli
operation $\sop{P}_i$ with the $i$th Clifford group operation
$\sop{C}_i$, to obtain a random Clifford group operations
$\sop{D}_i$~\footnote{This argument carries through independently of
  how the Clifford group operation is physically implemented (one
  pulse or multiple pulses, as long as $\sop{D}_i$ is applied
  consistently regardless of where may appear in a sequence).}. In
other words, a given sequence of Clifford group operations $\sop{C}_L
\sop{C}_{L-1} \cdots \sop{C}_2 \sop{C}_1$ becomes
\begin{align}
  & \underbrace{\sop{P}_{L+1} \sop{C}_L \sop{P}_{L}}_{\sop{D}_L}
    \underbrace{\sop{C}_{L-1} \sop{P}_{L-1}}_{\sop{D}_{L-1}}\cdots
    \underbrace{\sop{C}_2 \sop{P}_2}_{\sop{D}_2}
    \underbrace{\sop{C}_1 \sop{P}_1}_{\sop{D}_1}
\end{align}
which results in the randomized sequence of Clifford group operations
$\sop{D}_L \sop{D}_{L-1} \cdots \sop{D}_2 \sop{D}_1$. In essence, under PFR, a single realization of
a randomized sequence of Clifford group operations simply corresponds
to a different sequence of Clifford group operations.

It is possible to choose all $\sop{P}_i$ independently at random and
compensate for their action by flipping observed measurement outcomes
in post-processing (as, by construction, we only measure in the
computational basis). In order to simplify post-processing, we instead
choose $\sop{P}_{L+1}$ to cancel the effect that all other random
Pauli group operations would have on measurement results (i.e.,
$\sop{P}_{L+1}$ is a Pauli frame correction before measurement). In this way
the measurement outcome of the randomized and unrandomized experiments
can be treated exactly the same, with no additional post-processing
for the randomized experiments.

We can analyse the sequences above with the simplifying assumption of
gate-independent errors by replacing each operation with its noisy
counterpart. We write the noisy operations
$\tilde{\sop{D}}_i=\sop{E}\sop{D}_i$ (where $\sop{E}$ is an arbitrary
but fixed completely-positive trace-preserving (CPTP) map) to obtain
\begin{align}
 & \tilde{\sop{D}}_L \tilde{\sop{D}}_{L-1} \cdots \tilde{\sop{D}}_{2}
   \tilde{\sop{D}}_{1}, \\
=& ~\sop{E} \sop{D}_L \sop{E} \sop{D}_{L-1} \cdots \sop{E} \sop{D}_{2}
   \sop{E} \sop{D}_{1}, \\
=& ~\sop{E} \sop{P}_{L+1} \sop{C}_L \sop{P}_{L}
   \sop{E} \sop{C}_{L-1} \sop{P}_{L-1} \cdots
   \sop{E} \sop{C}_2 \sop{P}_2 \sop {E} \sop{C}_1
   \sop{P}_1.\label{eq:RandError}
\end{align}
Defining $\sop{P}^{\sop{C}}=\sop{C}\sop{P}\sop{C}^\dagger$, we can
write $\sop C\sop P = \sop P^{\sop C}\sop C$. Similarly, we define
$\sop P_{n:1}=\sop P_n\sop P_{n-1:1}^{\sop{C}_{n-1}}$ (with the base
case $\sop P_{1:1}=\sop P_1$). With these definitions, the entire
sequence can then be rewritten as
$  \sop{E}\,
  \sop{P}_{L+1:1}\,
  \sop{C}_{L}\,
  \sop{P}_{L-1:1}^{\sop C_{L-1}}\, \sop {E}\, \sop{P}_{L-1:1}^{\sop C_{L-1}}\,
  \sop{C}_{L-1}\,
  \cdots
  \sop{P}_{1:1}^{\sop C_1}\, \sop {E}\, \sop{P}_{1:1}^{\sop C_1}
  \sop{C}_1$,
where, in the experiments described here, we have chosen
$\sop{P}_{L+1:1}$ to be the identity. In other words, we
  choose $\sop{P}_i$ uniformly at random for $1\le i\le L$, and choose
  $\sop{P}_{L+1}$ to get a trivial $\sop{P}_{L+1:1}$.  Averaging over
many uniformly random choices of Pauli operations in Eq.~\ref{eq:RandError}, we transform each
$\sop{E}$ in the sequence into $\overline{\sop{E}}=\frac{1}{d^2}\sum_i
\sop{P}_i \sop{E} \sop{P}_i$, which correspond to twirling $\sop{E}$
over the Pauli group. This, in turn, ensures that the effective error
$\overline{\sop{E}}$ associated with each gate in the sequence
corresponds to a statistical mixture of Pauli
operations~\cite{DCEL09}, as desired~\footnote{Although the error in
  the last gate of the sequence is not randomized due to our choice to
  have the last randomization operation cancel all others, we can
  treat this as a measurement error. Alternatively, we could choose all
  randomization operations uniformly at random and changed the
  measurement outcome to undo the randomization, so that effectively
  the error in the last gate would also be twirled over the Pauli
  group.}.

The calculation outlined above does require rather strong assumptions
about the properties of the noise (i.e., that it is gate independent
and Markovian), but due to similarities to randomized benchmarking
(RB)~\cite{EAZ05,KLR+08,MGE11,MGE12,MGJ+12}, which has been shown to
require weaker assumptions~\cite{W+17}, we expect that these strong
assumptions are not strictly necessary. In the remainder of this paper
we focus on how to test such a hypothesis, and implement these test on
the natural imperfections of a superconducting qubit experiment.

\section{Hypothesis Testing~\label{sec:CV}}

The task of checking whether the result of applying PFR to an
experiment does indeed result in a Pauli channel is subtle. Modern
experiments have very high fidelity to ideal operations so checking
that the unrandomized errors are not well described by Pauli
channel---i.e., determining that PFR is necessary---is already
challenging, since error rates can be on the order of $10^{-3}$ or
less.  In both cases, it is natural to consider long sequences of
operations to amplify sensitivity to these small errors.

We choose to use long-sequence gate-set tomography
(GST)~\cite{BK+15,BGN+17,NGR+20} to observe these small effects, and
use a readily available open-source package for experiment design and
data analysis~\cite{PYGSTI}, with minor modifications. At heart, GST is a
sophisticated refinement of a quantum process
tomography~\cite{PCZ97,CN97}, providing a complete reconstruction of the action
of quantum operations. In particular, GST is an
iterative procedure that refines the tomographic reconstruction of a
set of gates by comparing predictions about long gate sequences to
experimental observations, and adjusting the reconstruction for better
agreement. Since long sequences allow for small perturbations to
accumulate, this technique yields unparalleled
accuracy~\cite{BK+15,Gre15,BGN+17,NGR+20}.

Even with a reconstruction in hand, another subtle question is how to
quantify the distance between reconstructed errors and a Pauli error
model---i.e., the degree of ``non-Pauliness'' of the noise. We use the
likelihood ratio test for this purpose~\cite{Neyman1933,Wilks1938},
which requires a hierarchy of nested models. The null hypothesis $H_0$
is taken to be that the statistics for each sequence in the GST
experiments leads to a separate Binomial distribution of
outcomes. More explicitly, for the null hypothesis we only assume that
the sequences correspond to reproducible experiments with well defined
measurement statistics, and ignore the gate structure of the
sequences. This corresponds to not making any assumption about
Markovianity or time independence of the system evolution. We then
consider two hypotheses nested within $H_0$: that each gate in the
sequence corresponds to a fixed linear operation acting on the system
(we call this first hypothesis $H_1$), and that each gate in the
sequence corresponds to a fixed Clifford group operation followed by a
fixed Pauli stochastic error operation (we call this second hypothesis
$H_2$). The consistency of $H_0$ with $H_2$, for some reconstructed
Pauli error model, will be taken as our measure of Pauliness.

As we indicated, these hypotheses are nested: $H_2$ is a special case
of $H_1$, and $H_1$ is a special case of $H_0$, meaning that if the
statistical tests indicated $H_2$ is consistent with $H_0$, the same
will be true of $H_1$. The statistical tests will only be
able to test if the proposed hypotheses are consistent with $H_0$, in
the sense that a hypothesis cannot have a higher likelihood than
another hypothesis it is nested into.

We fit data to a model under $H_0$ by maximum-likelihood estimation of
the Binomial distribution parameter $p$ associated with each GST
sequence. We fit data to a model under $H_1$ using progressive
refinement of maximum-likelihood estimation, a heuristic developed for
GST~\cite{PYGSTI}.  We fit data to a model under $H_2$ by projecting
the fit of $H_1$ into a generalized monomial matrix (described below),
determined by the corresponding noiseless Clifford group
operation. The first two fits are part of the standard routines within
GST, while the last fit is a small extension to the existing GST
routines.

The fitting of data to a model under $H_2$ proceeds as follows. In the
Pauli-Liouville
representation~\cite{Blum81,Leung00,Rahn2002,Fern2006,Chow2012,Kimmel2014,
  Johnson2015}, a Clifford group operation is a monomial matrix---each
row or column has a single non-zero matrix element, and this matrix
element is $\pm1$. In the presence of a Pauli error model, a noisy
Clifford group operation will be a generalized monomial matrix, where
the $\pm1$ elements of the noiseless matrix are replaced by numbers in
the interval $[-1,1]$ (but the 0 matrix elements remain unchanged).
Collectively, these matrix elements must live in a simplex equivalent
to the probability simplex for the Pauli channel~\cite{SMKE08}. Thus,
the projection of an $H_1$ model onto an $H_2$ model simply
corresponds to identifying which matrix elements should be set to zero
(i.e., which matrix elements are zero in the ideal gate), and then
adjusting the remaining non-zero matrix elements so that the resulting
matrix lies in the appropriate simplex.

\subsection{Badness-of-fit}

We quantify how well the data is explained with each of the hypotheses
discussed above by computing a metric for the quality of the fits
obtained. The basis for this calculation is ${\mathcal L}(H_i)$, the
likelihood of the observed data given the model fitted under a
particular hypothesis $H_i$.

Following Wilk's theorem~\cite{Wilks1938}, we know the log-likelihood
ratio $-2\log \frac{{\mathcal L}(H_i)}{{\mathcal L}(H_0)}$ has a
distribution that asymptotically (in the same size) approaches a
$\chi^2$ distribution with degrees of freedom given by the difference
in the dimensionality of the two nested hypotheses, under the
assumption that the null hypothesis is true. The mean and variance of
the asymptotic distribution for the log-likelihood ratio are
determined by the number of degrees of freedom. Given the fitted
models, the likelihood of the observations under the various
hypotheses are computed, and we follow the convention of reporting the
difference between the observed statistic and the mean predicted by
Wilk's theorem, in units of the standard deviation of the appropriate
$\chi^2$ distribution, and call this quantity $N_\sigma$. Intuitively,
if this ``badness-of-fit'' number is large, we favor the null
hypothesis (i.e., the hypothesis fit is bad), but if this number is
small, both the simpler hypothesis and the null hypotheses are valid,
and the simpler hypothesis is favored as a parsimonious model.  We
emphasize this ``badness-of-fit'' parameter cannot be obtained by
characterization techniques like RB that do not have an implicit model
built in.

The log-likelihood ratios allow us to quantify whether (a) the
observations are consistent with a Markovian error model (i.e.,
whether $H_1$ is plausible), and (b) whether the observations are
consistent with a Clifford group operation with a Pauli error model
(i.e., whether $H_2$ is plausible).  In particular, we are interested
in testing whether the answer to these questions changes when we apply
PFR to our experiments. For this, it is necessary to look at the
likelihood of hypotheses in different data sets (i.e., the
unrandomized and the randomized GST experiments). Likelihoods cannot
be meaningfully compared across different data sets. Instead, we
simply consider the plausibility of the different hypothesis for the
different data sets, while taking great care to ensure that the data
sets are representative of the same noise and error environment, as we
now describe.

\section{Experiments} \label{sec:experiments}

\subsection*{Device parameters}

To test the hypotheses of Section~\ref{sec:CV}, we implement the PFR
procedure on a superconducting qubit device.  The device consists of 
four fixed-frequency, transmon qubits,
designed to be similar to those described in~\cite{CGM+14}.  The
qubits are uncoupled but readout through a common Purcell filter.  For
the PFR experiment in this Letter, only one qubit (Q\textsubscript{1})
is measured. Q\textsubscript{1} is dispersively coupled to a readout
resonator with a center frequency of $\omega_r/2\pi = \, 7.112 \,
\rm{GHz}$, $\kappa/2\pi = \, 3.4 \, \rm{MHz}$, which is in turn
capacitively coupled to a quarter-wave Purcell filter with external $Q
= 22$ and a center frequency of $\omega_f = 7.27 \, \rm{GHz}$
\,\cite{JSM+14} enabling fast qubit readout.  Q\textsubscript{1} has a
fixed 0-1 transition frequency of $\omega_q/2\pi = \, 4.432
\,\rm{GHz}$ with an anharmonicity $\alpha/2\pi = \,$308 MHz.
Coherence times measured for Q\textsubscript{1} are $T_1 = 10 \, \mu
{\rm s}$, $T_2 = 13 \, \mu {\rm s}$ and a Hahn echo time of
$T_{\mathrm{echo}} = 16 \, \mu {\rm s}$ (the other qubits in the
device were not characterized). The pulses used were $50 {\rm ns}$
long, leading to an expected average gate infidelity (resp. diamond
norm distance) of at least $\sim0.2\%$ per pulse ($\sim1\%$ per
pulse).  Since we use 2 pulses in the implementation of the gates
discussed here, we expect an infidelity of no less than $\sim0.4\%$
and diamond norm distance of no less than $\sim2\%$ (ignoring all other
sources of error and ignoring the effects of PFR).

\begin{figure}
  \includegraphics[width=.45\textwidth]{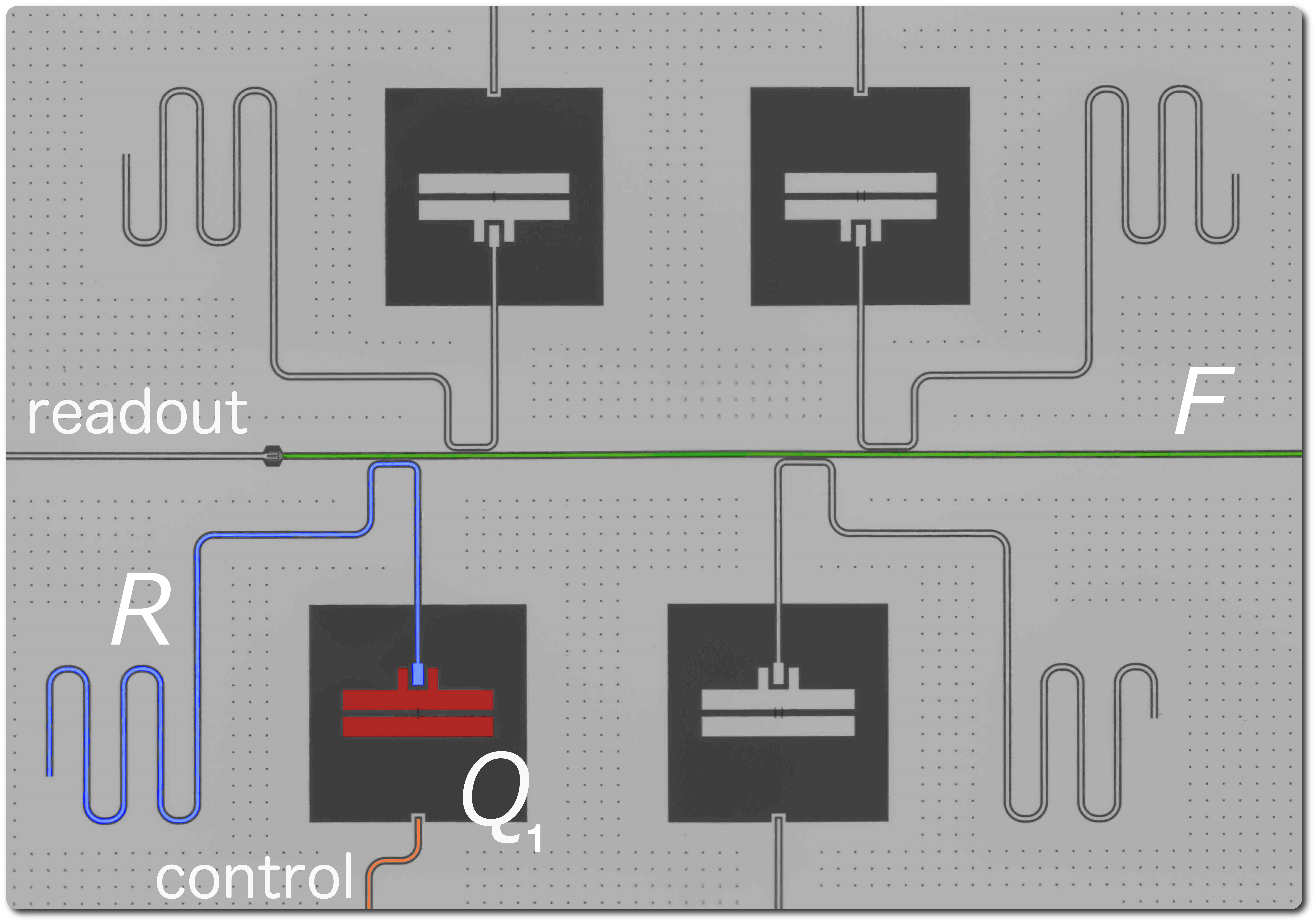}
  \caption{False-color micrograph showing qubit Q\textsubscript{1}
    (red), resonator {\it R} (blue) and Purcell filter {\it F}
    (green).  The qubit is dispersively coupled to a $\lambda/4$
    readout resonator which is capacitively coupled to a Purcell
    filter with a $Q = 22$.  Qubit control is done through a dedicated
    drive line (orange) and all qubit readout is done via a central
    feed line incorporating a Purcell filter.\label{fig:device}}
\end{figure}

\subsection*{Electronics and software stack\label{sec:hardware}}

Making the PFR process experimentally tractable requires leveraging a
complex software and hardware control infrastructure. The first hurdle
is the sheer number of experiments needed.  Long-sequence GST
($\ell$GST) experiments require a large set ($\sim$3500) of long
circuits, each with up to 6155 gates. To ensure high accuracy, we
produce a large number of measurement shots, 1000 per sequence.  Under
PFR, we take a single measurement shot per randomized sequence,
resulting in 1000 unique GST circuits for each of the $\sim$3500
$\ell$GST circuits originally specified. Thus, in total, we measure
over 3.5 million unique sequences to obtain high tomographic
reconstructions for the gates in the unrandomized and the randomized
experiments~\footnote{The number of randomizations could be much
  lower~\cite{WE16}, or, equivalently, more measurement shots could be
  collected per randomized sequences, but we chose to take the extreme
  limit of one shot per randomized sequence to avoid any subtle
  questions about how many shots would be safe to take per randomized
  sequence before correlations became significant.}. The second
challenge is running the experiments in a way that allows the most
direct comparison between the randomized and unrandomized cases---
doing so allows us to minimize the impact of drift when comparing how
the hypothesis tests from Sec.~\ref{sec:CV} fare on the different data
sets. To achieve this, the unrandomized and the randomized sequences
should be run in an interleaved fashion to ensure they experience the
same noise environment (to the extent possible). These requirements
necessitate hardware that can execute a large number of very long
circuits, and to quickly alternate between them.

To address these issues we use a custom sequence compiler written in
Julia~\cite{Julia} called {\tt QGL.jl}~\cite{BUQ+QGLJL}, providing a
$4\times$ compilation speed-up per circuit over an earlier Python
version through a combination of parallel computation and other
efficiency improvements---this ensured we were able to compile the 3.5
million unique sequences into pulse sequences in a reasonable amount
of time.  To minimize the runtime overhead we leverage a custom 
arbitrary pulse sequencer with gateware dedicated to implementing 
quantum circuits~\cite{BUQ+APS}.

A rough outline of the process is as follows. Standard, one-qubit
$\ell$GST sequences are created using the GST experiment and analysis
software {\tt pyGSTi}~\cite{PYGSTI}. For the data presented here, we
choose the maximal sequence length in GST to be 6150 gates, to ensure the
experiment will have high accuracy, and be sensitive to
non-Markovianity over timescales long compared to qubit coherence
times.  The $\ell$GST sequences are then randomized as
described in Sec.~\ref{sec:frame-rand}. It is worth emphasizing the
lengths of the randomized circuits are unchanged as the Pauli group
operations are combined with a neighboring Clifford group
operation~\footnote{The time necessary to generate these random
  sequences is minimal, on the order of a handful of minutes for all
  ~3.5 million unique sequences, since it requires only the simulation
  of Pauli errors propagating in Clifford circuits. If we were not
  taking the additional step of compensating for the randomization in
  the final measurement, the generation of these sequences would be
  even faster, since it would not require the propagation of the
  randomizing Pauli operations.}, much like the randomized compiling
proposal of Ref.~\cite{WE16}.

\begin{figure}
  \includegraphics[width=.48\textwidth]{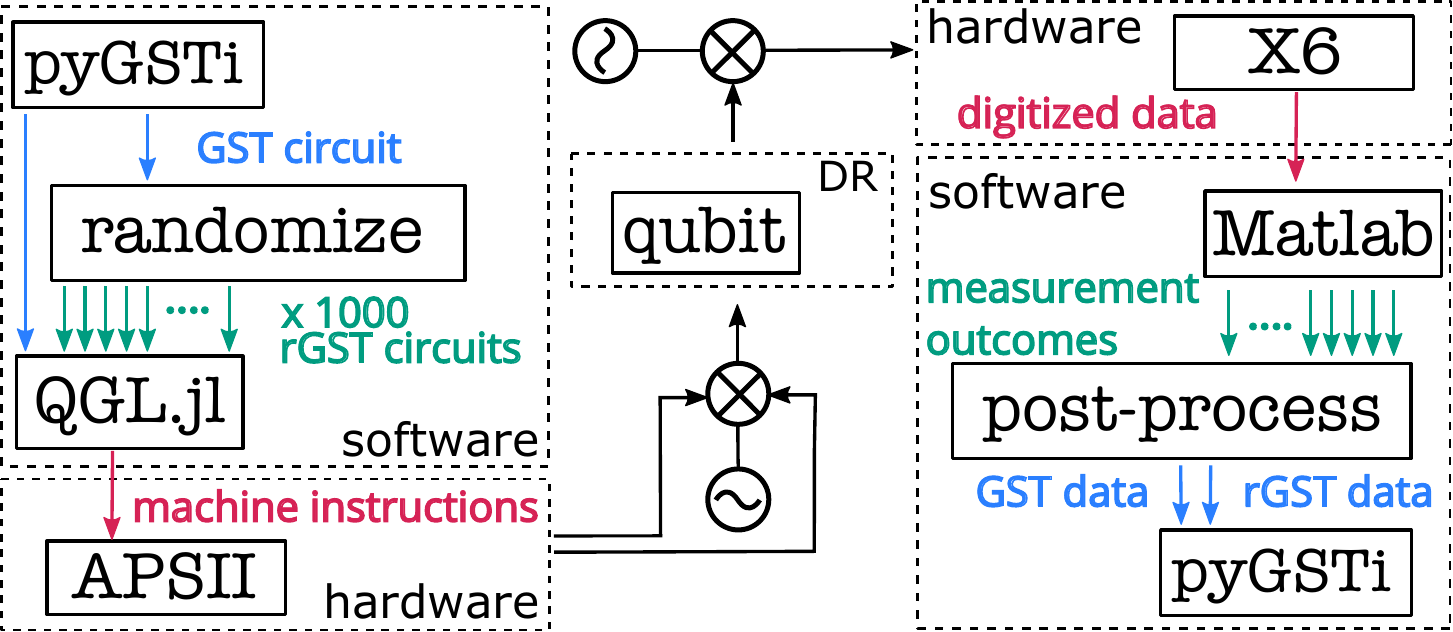}
  \caption{Experimental data flow.  Basic GST sequences are created
    using {\tt pyGSTi}~\cite{PYGSTI}. This basic experiment is then
    randomized 1000 times as described in
    section~\ref{sec:frame-rand}. Each randomization and the original
    experiment get compiled by the {\tt QGL.jl} compiler that
    translates sequence instruction into instructions implemented by
    APSII pulse sequencer~\cite{RJR+17}. The qubit response is digitized by
    an Innovative Integration X6 digitizer card and organized with a
    Matlab experimental framework. The single-shot data
    from each experiment is then post-processed into counts that {\tt
      pyGSTi} uses to reconstruct the gate set process matrices and
    the goodness-of-fit metrics.
    \label{fig:data_diagram}}
\end{figure}

The collection of uniquely randomized GST (rGST) circuits and the
original unrandomized circuits are then passed to the {\tt QGL.jl}
compiler which translates qubit gate instructions into machine
instructions.  This involves not only mapping high-level instructions
to control pulses but also time-ordering and synchronizing instruction
playback between all qubit control and readout channels. We note,
this process takes significantly more time than the generation or
randomization of the GST circuits. The compiled instructions are then
passed to the pulse sequencer which is used to control the qubit. Due to the
nature of the randomization process, the rGST experiments lack any
kind of repetition or subroutine structure which rules out any
efficient storage in hardware memory of a complete set of circuits. To
address this, the set of randomized experiments are broken into groups
of 10 in order to fit in the control hardware memory.  These 10 single shot runs of
rGST were then interleaved with 10 shots of unrandomized $\ell$GST.
The process is repeated 100 times for each data point in
Fig.~\ref{fig:reconstruction}. The complete process flow for a single
round of experiment generation is illustrated in
Fig.~\ref{fig:data_diagram}.

In the canonical construction of the Clifford group, elements are
composed of multiple native $\pi$ and $\pi/2$ pulses, which leads to
Clifford group elements being implemented by different numbers of
native gates/pulses non-uniform length.  To account for this, we use a
``diatomic'' implementation of the group where each Clifford group
operation is performed with two $X_{\pi/2}$ pulses of fixed length (50
ns) and three possible Z-frame updates~\cite{MWC+17,Johnson2015}. This
diatomic approach ensures all Clifford operations have equal duration.
The room temperature measurement signals were processed with an
autodyne technique described in Ref.~\cite{RJG+15} using the BBN-QDSP
digitization architecture~\cite{RJR+17} for the {\em Innovative
  Integrations X6-1000M} digitizer card. The final state assignment is
then fed into the {\tt pyGSTi} package for gate set reconstruction.
{\tt pyGSTi} also provides the likelihood of $H_0$ and $H_1$, while
custom code generates the likelihood of $H_2$---from these
likelihoods, we obtain the likelihood ratio statistic and
compare it to the predictions from Wilk's theorem.

\section{Results\label{sec:results}}

The experiment outlined in Sec.~\ref{sec:CV} was performed to test the
effectiveness of PFR.  This process was repeated seven times, each
taking roughly one hour to complete. The repetitions allow us to
observe how drift affects the results over an operationally meaningful
amount of time.

\begin{figure}
  \includegraphics[width=.45\textwidth]{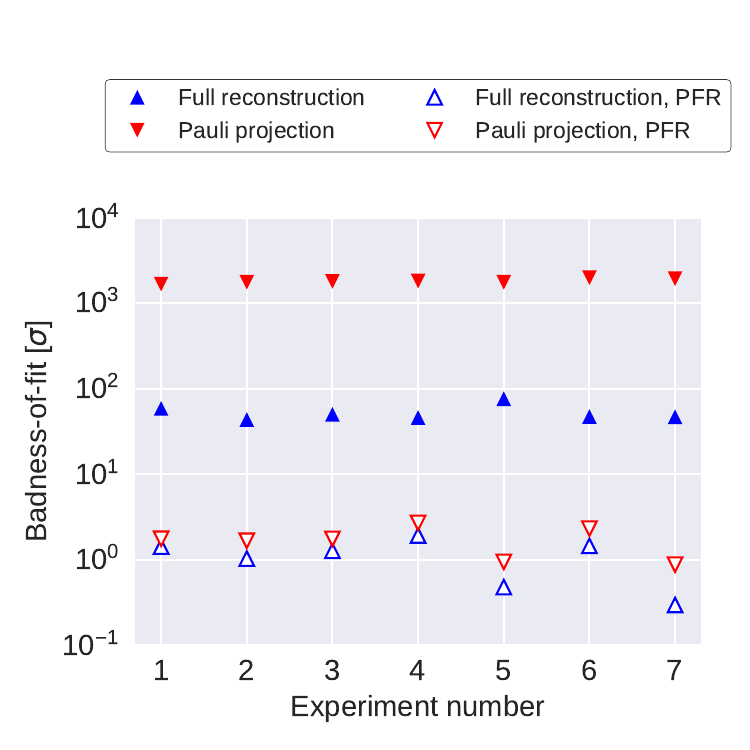}
  \caption{Badness-of-fit for the GST reconstructions under a
    Markovian error model ($H_1$, upward blue triangles), and a
    Markovian stochastic Pauli error model ($H_2$, downward red
    triangles), as quantified by the log-likelihood-ratio
    statistic. This statistic is presented as the difference from the
    predicted mean of the $\chi^2$ distribution (from Wilk's theorem),
    in units of the standard deviation of that same distribution,
    under the assumption that $H_0$ is true. Both experiments without
    randomization (full triangles) and with randomization (empty
    triangles) are considered.  As discussed in the text, the ~3.5
    million unique sequences that comprise the tomography experiments
    for randomized and unrandomized experiments are measured in an
    interleaved fashion, so that both reconstructions should
    experience the same physical noise conditions. The entire
    collection of tomography experiments are repeated 7 times to
    illustrate the behavior observed persists, and thus unlikely
    to be the result of statistical fluctuations, and is robust to
    drift in our system (each of these 7 experiment lasted
    roughly one hour).
    \label{fig:reconstruction}}
\end{figure}

One of the critical questions of this work is the validity of $H_2$
(the hypothesis that gates are well described by Clifford group
operations followed by stochastic Pauli noise) and the Markovian
behavior of qubit evolution under PFR. Data addressing this question
can be seen in Fig. \ref{fig:reconstruction} where the GST model
violation is plotted in terms of $N_{\sigma}$ both with and without
randomization.  Several features are immediately apparent: (1) the
Markovian fits ($H_1$) to the unrandomized experiments (filled
triangles) are orders of magnitude worse than the randomized
experiments (empty triangles), (2) the data projected to a Pauli error
model ($H_2$) in the unrandomized cases (red filled triangles
pointing down) is roughly three orders of magnitude worse than the
randomized experiments (red empty triangles pointing down), (3) there
is little difference between the quality of the fits under all the
hypotheses for the randomized experiments (empty triangles). In terms
of the hypotheses outlined previously, for unrandomized experiments
there is a large likelihood discrepancy between $H_0$ and the simpler
hypotheses, greatly favoring the non-Pauli, non-Markovian $H_0$ model
($H_1$ is $43\sigma$ to $76\sigma$ away from the predictions from
Wilk's theorem, and $H_2$ is $1754\sigma$ to $1987\sigma$ away), while
for the randomized experiments all hypotheses have comparable
likelihood (within $0.3\sigma$ to $2.7\sigma$ of the predictions from
Wilk's theorem), so it is reasonable to take the simplest hypothesis
(the Markovian, stochastic Pauli error model $H_2$) as the best
explanation for those observations.

We should note that, despite the base level of $H_1$ model violation
measured in the unrandomized data (a signature of non-Markovianity)
appearing large at $1987\sigma$, it is largely consistent with
observations in other systems under similar circumstances (see, e.g.,
ion trap experiments without drift control or decoupling
pulses~\cite{BGN+17}).

\begin{figure*}
  \begin{center}
  \includegraphics[width=0.95\textwidth]{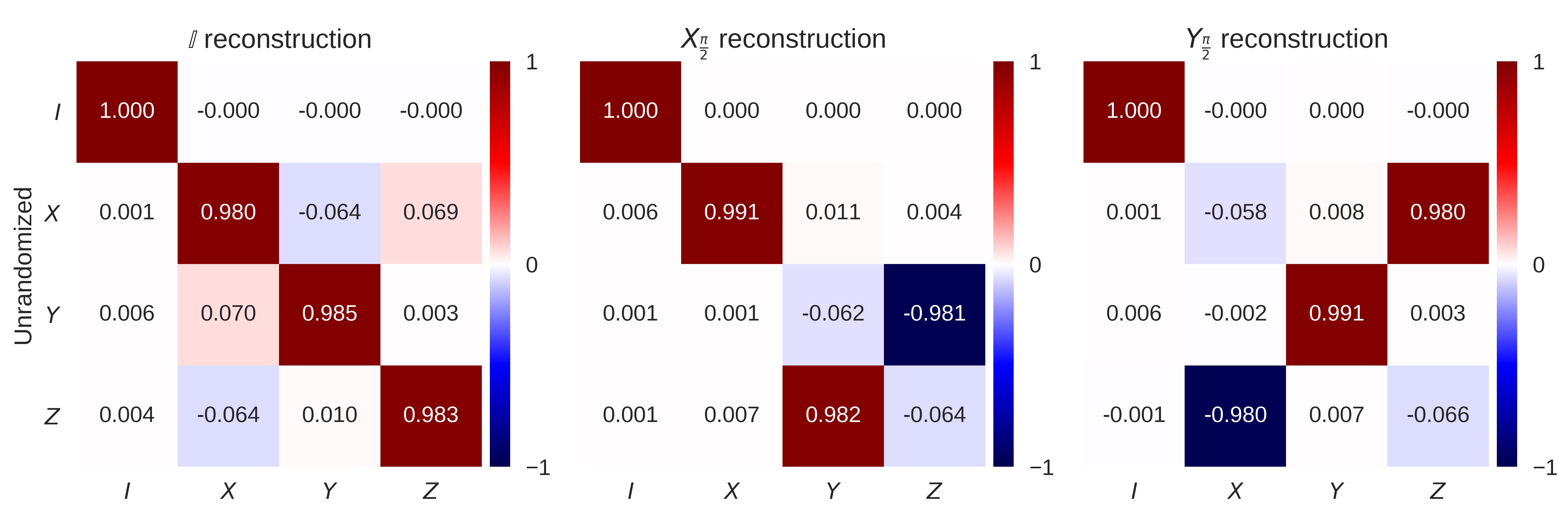}
  \includegraphics[width=0.95\textwidth]{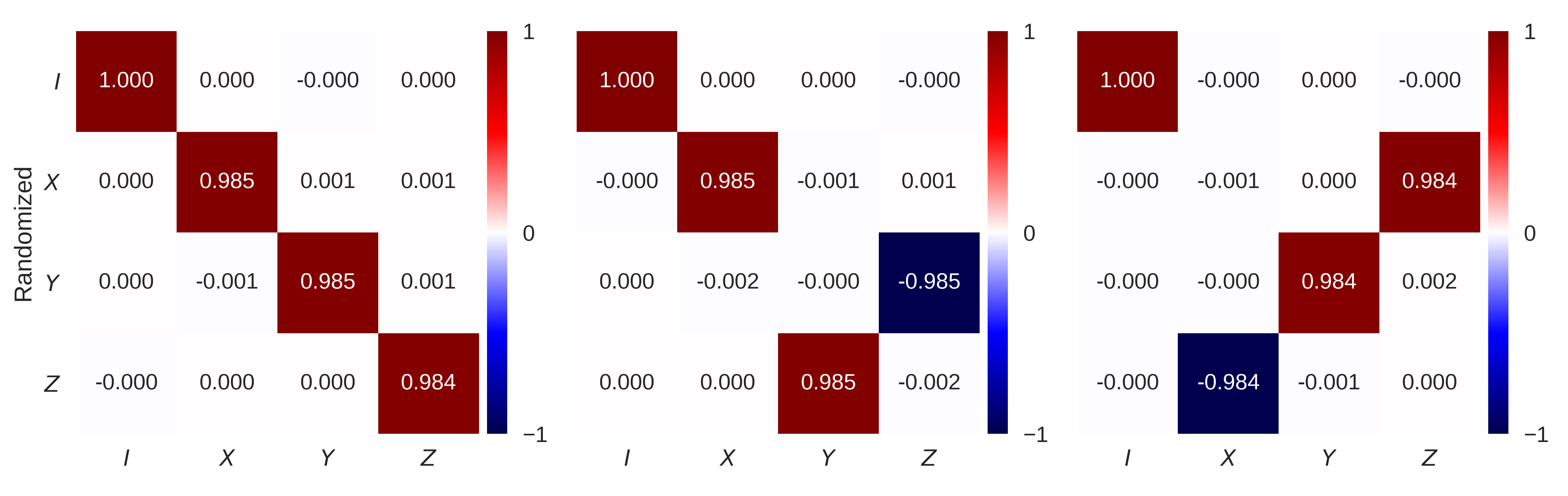}
  \end{center}
  \caption{Matrix representations of the reconstructed processes
    corresponding to the $\mathbb{I}, X_{\frac{\pi}{2}}$, and
    $Y_{\frac{\pi}{2}}$ operations in the first of the seven
    experiments performed (details of the other six experiments and
    confidence interval computation can be found in
    Ref.~\cite{WS2020}). Without randomization (top row) there
    are significant off-diagonal contributions, corresponding to
    non-Pauli errors. With randomization (bottom row) there are no
    statistically significant off-diagonal contributions, indicating
    the errors correspond to a Pauli error model, as expected. Error
    bars (95\% confidence) for this experiment are smaller than
    $\pm0.0028$ (unrandomized) and $\pm0.0023$ (randomized).
\label{fig:matrix}}
\end{figure*}

These features strongly indicate the noise in the absence of
randomization is not well described by a Markovian error model, which
follows from comment (1) above.  Also apparent from comment (2) is
that even the best Markovian error model is not well approximated by a
Pauli model in the absence of randomization. Conversely, these
features indicate the noise under PFR is very well described by a
Markovian Pauli error model.  In much simpler terms, the features of
non-Pauli error models (i.e., non-trivial off-diagonal matrix elements
in the Pauli-Liouville representation~\cite{KSR+14}) are insignificant
in the reconstructions of the randomized experiments, as
Fig.~\ref{fig:matrix} illustrates.  These separations are persistent
over many repeats of the experiment, and the separation of many orders
of magnitude indicates that PFR worked in these experiments not only
quantitatively, but qualitatively, i.e., unradomized experiments have
strong non-Markovian features, while randomized experiments were well
explained by Markovian Pauli error models.

\subsection*{Behavior of error metrics under randomization\label{app:error-metrics}}

The badness-of-fit results illustrate that the models derived under
randomization are much more useful in explaining the observations than
the models derived without randomization. A natural question that
arises is whether this comes at the cost of degrading the performance
of the gates. Here we demonstrate that this is not the case, and that
in fact the performance of the gates improves under PFR.

\begin{figure}
  \includegraphics[width=\columnwidth]{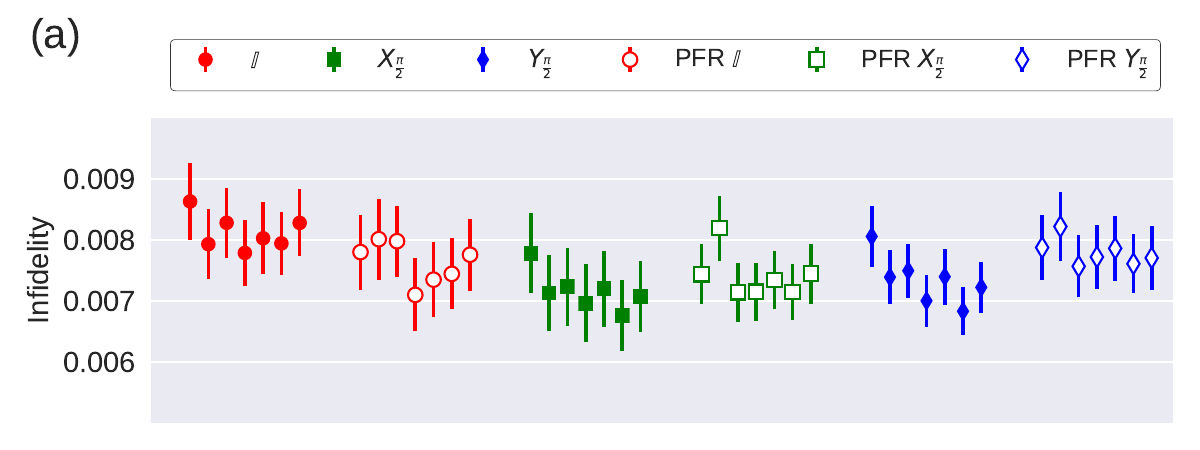}
  \includegraphics[width=\columnwidth]{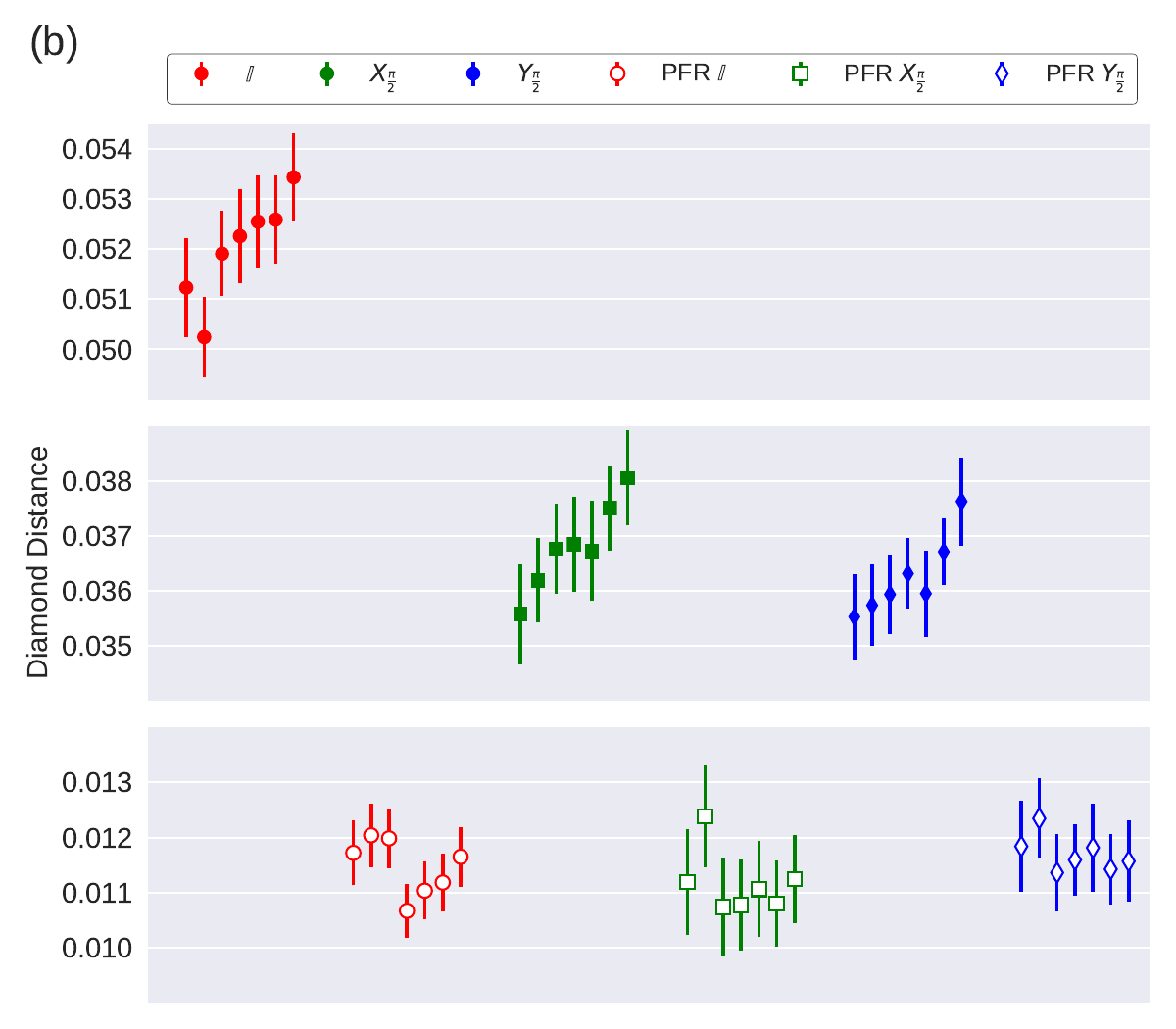}
  \caption{(a) Average gate infidelity and (b) diamond norm distance
    estimates for all three gates (colors) on a sequence of seven
    separate experiments over several hours. Data for different gates
    is horizontally offset for clarity, but we emphasize each of the
    experiments leads to a reconstruction of all three gates, for both
    randomized (full symbols) and unrandomized (empty symbols)
    gate-sets under identical conditions---the only distinction
    between the seven experiments is the passage of time.
    All quantities are computed from the reconstructed process
    matrices.  For the randomized case (empty symbols), the infidelity
    and the diamond norm are comparable, at $\approx1\%$.  For the
    unrandomized experiments, there is significant deviation between
    the diamond norm error rate and the infidelity, suggesting the
    presence of coherent errors that affect the infidelity metric only
    weakly (and which are suppressed in the randomized experiments).  A
    monotonic upward trend in the diamond norm distance of
    unrandomized experiments (full symbols) implies the presence of
    systematic drift in the control pulses, which is also suppressed by
    randomization (empty symbols). Error bars are $95\%$ confidence
    intervals (analytically for (a), and with the Hessian provided by
	pyGSTi for (b)).
    \label{fig:infidelity-diamond-norm}}
\end{figure}

We computed the average gate infidelity~\footnote{In this work we use
  the standard definitions of average gate infidelity as $1 -
  \mathcal{F}(\sop{U},\tilde{\sop{U}})$, with average gate fidelity
  $\mathcal{F}$ defined as~\cite{Nielsen02}
\begin{align*}
  \mathcal{F}(\sop{U},\tilde{\sop{U}}) = \frac{\tr~\tilde{\sop{U}}\sop{U}^\dagger + d}{d^2+d}
\end{align*}
} and the diamond norm distance~\footnote{The diamond distance between
  two processes~\cite{KSV02,W+09} is defined as
\begin{align*}
  \mathcal{D}_{\diamond}
  = \frac{1}{2}||\tilde{\sop{U}} - \sop{U} ||_{\diamond}
  = \mathop{{\rm sup}}_{\rho}\frac{1}{2} ||(\tilde{\sop{U}} \otimes \sop{I}_d - \sop{U} \otimes \sop{I}_d) (\rho) ||_1,
\end{align*}
with $d$ being the total system dimension.}  for the reconstructed
gates under normal operation and under PFR, as depicted in
Fig.~\ref{fig:infidelity-diamond-norm}. The observed average gate
infidelities (resp. diamond norm distances) are roughly double
(quadruple) the expected coherence limits of the device, which may be
explained by dynamical effects in the gate implementation which may be
addressed by more careful pulse shaping~\cite{Chow2010} (and which are
not accounted for in the coherence limit calculation mentioned
earlier). We observe no appreciable difference between the infidelity
of randomized and unrandomized experiments, while the diamond norm
distance is reduced by a factor of 3-5 under PFR. This is consistent
with the well known behavior of the infidelity and the diamond norm
under small coherent errors---namely, the infidelity is only sensitive
to coherent errors to second order, while the diamond distance is
sensitive to first order~\cite{KLDF2016}.

The diamond norm distance of unrandomized experiments monotonically
increases over the course of the 7 experiments, a behavior consistent
with drift in the qubit and control parameters with respect to
calibrations. The qubit control parameters are calibrated only once,
at the begining of the first experiment. This drift may at least
partially explain the violation of the time-independent Markovian
model represented by $H_1$, since these parameters appear to be
continuously and systematically drifting throughout the 7 experimental
runs, but this is a small effect since $H_1$ still makes accurate
predictions within each of the runs.

It should be noted that the drift is not apparent in the randomized
experiments (even in the diamond norm distance), despite these
experiments being run under the same conditions as the unrandomized
experiments. This indicates that the drift was averaged away under PFR,
a behavior consistent with coherent errors.

\subsection*{Access to data}

The experimental data, along with scripts used to perform the analysis
and plot the results, can be found in Ref.~\cite{WS2020}. These
include the full tomographic reconstruction of gatesets for all 7
experiments, along with the raw counts needed for these reconstructions.

\section{Summary}

We have demonstrated that Pauli-frame randomization reduces both the
non-Markovian features and the non-Pauli model features of errors in
single qubit experiments. This demonstration relies on long-sequence
gate-set tomography, which yields high accuracy reconstructions of all
operations used in the experiments.  This in turn required a
high-degree of automation to capture and process the $\sim7$ million
measurement shots/hour. In the absence of randomization, the
experiments were shown to have strong non-Markovian features, and the
best Markovian model in that case was also shown to have strong
features inconsistent with Pauli error models.  In the presence of
Pauli-frame randomization, the experiments were shown to be highly
consistent with a Markovian Pauli error model, as predicted.  As
quantified by log-likelihood-ratio statistic, the violation of
Markovian and Pauli error models in the unrandomized experiments is
highly-significant, as high as $1987\sigma$, while the violation of
Markovian Pauli error models in the randomized experiments are
statistically insignificant, less than $2.7\sigma$ in most of the
experiments.  This several orders-of-magnitude separation between
randomized and unrandomized experiments was persistent across seven
repeats of the experiment, indicating the noise-shaping effect of
Pauli-frame randomization is robust to drift in the control parameters
and fluctuations in the noise environment.

Areas for future work include speeding up the experiments using
techniques such as active reset~\cite{RBO+17,RBL+12}, and pushing
randomization process onto the hardware FPGA, which would allow for
data acquisition of randomized Clifford group circuits without the
user having to manually pre-compile random circuits.

After this work was completed similar results were independently
reported by Hashim et al.~\cite{HNM+20}.

\section{Acknowledgements}

We thank Robin Blume-Kohout, Erik Nielsen, Joel Wallman, and Joseph
Emerson for fruitful discussions, and the anonymous referees for
comments that helped improved the presentation of the paper. We also
thank Alan Howsare, Adam Moldawer and Ram Chelakara at Raytheon
Integrated Defense Systems for sample fabrication.

The qubit design and fabrication were funded by Internal Research and
Development at Raytheon BBN Technologies and directed by Thomas Ohki.
This work was funded by LPS/ARO grant W911NF-14-C-0048.  The content
of this paper does not necessarily reflect the position or the policy
of the Government, and no official endorsement should be inferred.

\section*{Contributions}
MW, GR and MPS wrote the manuscript. MW, GR, and DR performed the
experiment, while CR and BJ built the control software and electronics
infrastructure.  GR and MW designed the device and assisted in
fabrication.  MPS proposed the experiment, and wrote the experiment generation
scripts. MW and MPS performed the data analysis.

\bibliography{References}

\end{document}